\documentclass[12pt]{article}
\usepackage[dvips]{graphics}
\usepackage{graphicx}
\usepackage{dcolumn}
\usepackage{bm}
\usepackage{amsmath}
\usepackage{amssymb}
\usepackage{color}
\definecolor{darkblue}{rgb}{0,0,.5}
\definecolor{darkgreen}{rgb}{0,0.5,0}
\definecolor{darkred}{rgb}{0.5,0,0}
\usepackage[colorlinks = true,
            linkcolor = darkblue,
            urlcolor  = darkblue,
            citecolor = darkblue,
            anchorcolor = blue]{hyperref}

\newcommand{\beq}{\begin{equation}}
\newcommand{\eeq}{\end{equation}}
\newcommand{\beqa}{\begin{eqnarray}}
\newcommand{\eeqa}{\end{eqnarray}}

\begin{document}
\def\ii{\'\i}

\title{Vacuum fluctuation inside a star and their consequences for
neutron stars, a simple model}

\author{
G. Caspar$^1$, I. Rodr\ii guez$^1$, Peter O. Hess$^2$ and Walter Greiner$^1$ \\
{\small\it
$^1$Frankfurt Institute for Advanced Studies, Wolfgang Goethe University,}\\
{\small\it 
Ruth-Moufang-Strasse 1, 60438 Frankfurt am Main, Germany}\\
{\small\it
$^2$Instituto de Ciencias Nucleares, UNAM, Circuito Exterior,}\\
{\small\it
 C.U., A.P. 70-543, 04510, Mexico D.F., Mexico}
}

%\date{\bf Version under development}

%\twocolumn[
%\begin{@twocolumnfalse}
\maketitle
\begin{abstract}
Applying semi-classical Quantum Mechanics, the vacuum fluctuations within a star are determined, assuming a constant mass density and applying a monopole approximation. It is found that the density for the vacuum fluctuations does 
not only depend linearly on the mass density, as assumed in a former publication, where neutron stars up to 6 solar masses were obtained.
This is used to propose a simple model on the dependence of the dark energy to the mass density, as a function of the radial
distance $r$.
It is shown that stars with up to 200 solar masses 
can, in principle, be obtained. Though, we use a 
simple model, it shows that in the presence of vacuum fluctuations
stars with large masses can be stabilized and probably stars up to any mass can exist, which usually are identified as black holes. 
\\
\end{abstract}
%\end{@twocolumnfalse}

\section{Introduction}

The theory of General Relativity (GR) \cite{gravitation} has
passed up to now all observational 
%%%new-start
%and experimental
%%%new-end
tests. 
However, all these tests are for weak gravitational fields \cite{tests-GR} (to which we also count
the gravitational field in the Hulse-Taylor pulsar \cite{hulse-taylor}), compared to the field strength near the 
Schwarzschild radius. 

The standard GR predicts singularities, due to the black-hole solution.
This solution has a singularity in the center and
%%%new-start
%,
%for non-rotating stars,
%%%new-end
a coordinate singularity 
%%%new-start
%at the Schwarzschild radius, 
%%%new-end
called the
{\it event horizon}. No information below this event horizon can reach an external
observer, even nearby.

In our philosophical understanding, {\it no theory should have a singularity}, not 
even a coordinate singularity of the type discussed above. The appearance of a singularity hints to the incompleteness of a theory.
While this is generally accepted
for the singularity at the center, the discussion is going on whether this also applies for the coordinate singularity.

In recent publications \cite{hess2009,schoenenbach2012} an algebraic extension of GR to pseudo-complex (pc) coordinates was proposed, denoted as {\it pseudo-complex
General Relativity} (pc-GR). The pc-coordinates have
the structure $X^\mu = x^\mu + I y^\mu$, with $I^2=1$. Due to the property of $I^2=1$, variables with 
%%%new-start
%$x^\mu = y^\mu$
$x^\mu = \pm y^\mu$
%%%new-end 
have no inverse, thus 
the pc-variables form not a field but a ring. The 
absolute value squared $\mid X^\mu \mid$ = $X^{\mu *} X^\mu$ 
of these variables, with $X^{\mu *}=x^\mu -Iy^\mu$,
are zero, though the variables themselves are different from zero. A modified
variational principle was proposed, which states that the variation of the action,
which is now pseudo-complex, has to be proportional to a function whose absolute value
is zero, called also a {\it generalized zero}. As a consequence, the 
Einstein equations change and,
%%%new-start
in the limit of $y^\mu = 0$,
%%%new-end
have the form

\beqa
G_{\mu\nu} & = & -8\pi T^\Lambda_{\mu\nu}
~~~,
\label{intro-1}
\eeqa
where the the gravitational constant and the velocity of light were set to 1.
The $T^\Lambda_{\mu\nu}$ is the energy momentum tensor of the dark energy and the index $\Lambda$ denotes this association.
%%%new-start
The same equation is obtained, adding in the standard GR
by hand an energy-momentum tensor on the right hand side
of the Einstein equations. 
%%%new-end

Using semi-classical Quantum Mechanics (QM), where the background metric is fixed, it was shown that a mass changes the vacuum structure 
in its vicinity. The 
fluctuations increase toward the Schwarzschild radius 
and become
infinite there \cite{visser-boul}. 

Because a quantized theory of gravitation does not exist yet,
one relies on semi-classical QM, which does not permit to include the back-reaction of the vacuum fluctuation in a consistent way and is not applicable for strong
gravitational fields. 
Therefore, we proceeded to assume the presence of vacuum fluctuations, proposing a radial dependence such that 
for large distances, compared to the Schwarzschild radius, the corrections to standard GR can not be measured yet. For the dark energy a fluid model was 
chosen, leading to 
corrections of the metric in the order of  $1/r^3$.
(In \cite{MNRAS2013} other radial dependencies were investigated with similar
qualitative results). 
This did permit to include the back-reaction to the metric
and to solve the Einstein equations. 

The procedure reflects a general principle, namely: {\it Mass not only curves the space but also modifies the vacuum structure, such that the metric itself changes}.
Because we used a fluid model, an additional parameter $B=bm^3$ was introduced, where $m$ is the mass in units of length and $b$ is the actual parameter,
measuring the coupling between the central mass to the vacuum fluctuations outside
the star. By
an adequate choice of this parameter, namely that $g_{00}$ 
%%%new-start
%=
satisfies
%%%new-end
$\left(1-\frac{2m}{r}+\frac{B}{2r^3}\right)$ $>$ 0, the $b$ obtains a lower
bound, such that there is {\it no event horizon}!

In \cite{MNRAS2013} the consequences for the case of the Kerr-solution 
were investigated.
It was shown that the orbital frequency of particles in a circular orbit
exhibit a maximum in the frequency which from there on is decreasing again toward 
lower radial distances. 

Observing {\it Quasi Periodic Objects} (QPO) in the accretion disk of galactic black holes
\cite{qpo1,qpo2,qpo3,qpo4}, assuming that they correspond to
local bright spots in the disk following its orbital motion, one can deduce a radial distance using the prediction of the theory (GR or pc-GR). Measuring at the
same time the redshift, using $K-\alpha$ lines, also has as a result a given
radial distance. For a consistent theory, in both measurements the 
deduced radial distances have to be the same.
{\it And here is the problem}: They agree in pc-GR but not in GR
\cite{boller-bop}. There is still a discussion going on if the QPO follows the
orbital motion of the disk or are due to oscillations within the disk \cite{oscl} provoked by the stellar partner,
which would change the conclusions. Our argument that the QPO's in accretions
disks around galactic black holes correspond to local bright spots, following the motion of the disk, 
is based on their observation 
in the accretion disks around large masses in the center
of galaxies, with no stellar partner nearby, and that the physics in both should be very similar.    

Another prediction of pc-GR is that the accretion disk appears brighter and
exhibits a dark ring \cite{MNRAS2014}.
A model of a thin, infinitely extended disk \cite{pagethorne} was applied.
This dark ring is due to the maximum
of the orbital frequency of particles in a circular orbit. At the maximum 
shear forces between particles in two neighboring orbits is small and thus little
heat is additionally created.

%%%new-start
As one notes, pc-GR provides definite observable predictions,
which will be observed in near future.
%%%new-end

A further application of pc-GR is given in 
\cite{rodriquez2014a},  where the question was investigated if due to the presence of dark energy, neutron stars with larger
masses than 2-3 solar masses are possible. Due to the lack of knowledge
on how the dark energy couples to the mass distribution
within a star, a simple ansatz
was used, namely that the dark energy density $\epsilon_\Lambda$ is
proportional to the mass density $\epsilon_m$. The proposed coupling reads
$\epsilon_\Lambda = \alpha \epsilon_m$, with $\alpha < 0$. As a result
neutron stars up to 6 solar masses were obtained and their stability
was proven. Larger masses could not be realized, because
the coupling turned out to be too strong near the surface. We attributed it
to the proportionality between the two densities.

In order to get a better estimate of the coupling between the dark energy and the
mass distribution, we are lead to consider semi-classical QM with mass present.
%%%new-start
It refers to the interior of normal stars and 
known neutron stars.
%%%new-end
The suggested coupling will be treated as a phenomenological model.

The outline of the paper is as follows. In Section 2 we present the main points and
formulas in the semi-classical treatment, first outside a star and then
within a star, using the monopole approximation which permits a simpler treatment. The Schwarzschild metric is considered as the background, i.e.,
we discuss non-rotating stars. Two different kinds of mass distributions are considered. The first is a constant mass density within the star
and the second one is a density
which falls off such that it simulates realistic density distributions, except near the surface of a star. 
We will show that the coupling between the
dark energy and mass density
%%%new-start
in the interior region
%%%new-end
has to diminish toward the radius of the star,
even for a constant mass density.
The models discussed are simplistic but reveal the 
key properties.

In the same section,
a simple radial dependence of the dark energy 
%%%new-start
%density 
%%%new-end
to
the mass density is proposed and 
%%%new-start
%which will be 
%%%new-end
used in Section 3 to calculate the
masses for the stars. We show that now
up to 200 solar masses are possible.
Higher masses could not be obtained due to numerical difficulties. 
The result indicates that it is a matter of the correct coupling that 
"neutron stars" of arbitrary mass can be created, or in different words, that the
large masses observed in the center of nearly every galaxy are "neutron stars", 
which are very black and rather simulate a black hole. The main motivation is to
give a {\it prove of principle} that 
stars can acquire arbitrarily large masses, without
forming a black hole. (Though, it is doubtful that such objects
still resemble the neutron stars as we know them.) 

In Section 4 conclusions are drawn. 

\section{Semi-classical Quantum Mechanics and Vacuum fluctuations}

A very good introduction to semi-classical Quantum Mechanics is given in
\cite{birrell}. The main goal is to determine the vacuum fluctuations in a fixed
background metric, for example a Schwarzschild metric.
The physical quantity of interest is the expectation value of the energy-momentum
tensor $T_{\mu\nu}$, which is a local quantity and is 
preferred versus the observation of particle numbers.
The measure of particle numbers is observer 
dependent, while $T_{\mu\nu}$ transforms under
a relativistic transformation, relating one system to another equivalent one in a well defined way.
The energy-momentum tensor has to be 
%%%new-start
regularized and
%%%new-end
renormalized,
%%%new-start 
%and
where 
%%%new-end
the methods are
explicitly presented in \cite{birrell}.

The 
%%%new-start
%renormalization 
regularization/renormalization 
%%%new-end
of the energy-momentum tensor is still quite involved
and approximate methods have to be applied in order to determine its
expectation value in 4-space-time \cite{visser-boul,page}. A simpler approach
is followed invoking the {\it monopole approximation} \cite{birrell,padman}. The
monopole approximation consists of assuming a spherical symmetry and restricting 
the length element squared to the time and radial component, i.e.

\beqa
ds^2 & = & e^\nu dt^2 - e^\lambda dr^2
~~~,
\label{eq-1}
\eeqa
with $g_{00}=e^\nu$ and $g_{11}=e^\lambda$.
Defining the tortoise coordinate \cite{padman} via

\beqa
d\xi^2 & = & e^{\lambda - \nu}dr^2
\label{eq-2}
\eeqa 
leads to

\beqa
ds^2 & = & C(r) \left( dt^2 - d\xi^2\right)
~~~,
\label{eq-3}
\eeqa
with $C(r)=e^\nu = g_{00}$. Because we study a time-independent spherical metric,
the factor $C$ depends on the radial distance only.

Defining further

\beqa
x^+=t+\xi & , & x^-=t-\xi
~\rightarrow~ t=\frac{1}{2}\left(x^++x^-\right) ~,~ 
\xi =\frac{1}{2}\left(x^+-x^-\right)
~~~,
\label{eq-4}
\eeqa
leads to the line element

\beqa
ds^2 & = & C(r) dx^+dx^-
~~~.
\label{eq-5}
\eeqa

In this 2-dimensional space the expectation value of $T_{\mu\nu}^{2D}$ can be 
determined readily \cite{birrell,padman}:

\beqa
\langle T_{\pm\pm}^{2D}\rangle & = & -\frac{1}{48\pi} C^{\frac{1}{2}} 
\partial_\xi^2 c^{-\frac{1}{2}}
\nonumber \\
\langle T_{\pm\mp}^{2D} \rangle & = & \frac{C}{96\pi} R ~,~ 
(R=- C^{-1}\partial_\xi^2 {\rm ln}C)
~~~.
\label{eq-6}
\eeqa

Transforming the coordinates $x_\pm$ to $t$ and $r$, we
obtain the alternative expression

\beqa
 \left \langle T_{t}^{t~2D} \right \rangle & = & 
- \frac{1}{384\pi} e^{-\lambda} \left ( \nu^\prime \right )^2 
\nonumber \\
 \left \langle T_{r}^{r~2D} \right \rangle & = & 
-\frac{e^{-\lambda}}{192\pi} 
\left ( 2\nu^{\prime\prime} - \frac{1}{2} \left ( \nu^\prime \right )^2 
- \nu^\prime \lambda^\prime \right )
\nonumber \\
\left \langle T_r^t \right \rangle & = &
\left \langle T_t^r \right \rangle ~=~ 0 
~~~.
\label{eq-6a}
\eeqa
Using the Tolman-Oppenheimer-Volkov (TOV) equation \cite{adler}, which relates
$\nu$, $\lambda$ and their derivatives in $r$ to the density, pressure and the 
derivative of the pressure, we obtain

\beqa
 \langle T_{t}^{t~2D} \rangle & = & - \frac{1}{96\pi} 
\frac{\left ( m + 4\pi p_r r^3 \right )^2}{r^3\left (r-2m \right )} 
\nonumber \\
 \langle T_{r}^{r~2D} \rangle & = & - \frac{1}{96\pi} 
\left[  16\pi \left ( \varepsilon_m+ p_r^\prime r + 3 p_r \right ) 
\right.
\nonumber \\
&& \left.
- \left( 8r - 10 m -4m^\prime r + 8\pi p_r r^3 \right)
  \frac{m + 4\pi p_r r^3}{r^3\left (r -2m \right )}\right]
	~~~.
\label{eq-6b}
\eeqa

The 4-dimensional result is approximated by

\beqa
\langle T_{\mu\nu}^{4D}\rangle & = & \frac{1}{4\pi r^2}
\langle T_{\mu\nu}^{2D}\rangle
~~~,
\label{eq-7}
\eeqa
with $T_{\mu\nu}^{4D}=T_{\mu\nu}$. {\it This is 
the monopole approximation}.
The advantage of it is its simplicity, but deviations
from the exact result are to be expected.

For the Schwarzschild solution, in \cite{padman} the expectation value of the energy-momentum tensor components
%%%new-start
in the exterior region of a star
%%%new-end
was determined in two dimensions. After the
transformation to the coordinates $t$ and $r$, one obtains for the
4-dimensional expectation values outside the mass distribution,
defining $r_s=2m$ (the  Schwarzschild radius) and skipping the index "4D",

\beqa
\langle T_t^t \rangle & = & \varepsilon_\Lambda ~=~
-\frac{1}{1536\pi^2 r_s^4} \left ( \frac{r_s}{r} \right )^6 \frac{1}{1 - \frac{r_s}{r}}
\nonumber \\ 
\langle T_r^r \rangle & = & p^\Lambda_r ~=~
\frac{1}{96\pi^2 r_s^4} \left ( \frac{r_s}{r} \right )^5 \frac{1 - \frac{3}{8} \frac{r_s}{r}}{1 - \frac{r_s}{r}}
~~~,
\label{eq-8}
\eeqa
where we have transformed to the 
mixed components of the tensor (one component below and the
other above), which permit
a direct comparison to the density, e.g., $T_t^t =\varepsilon_\Lambda$. The
index $\Lambda$ refers to its interpretation as dark energy. The result agrees in
structure with \cite{visser-boul}, though the factors are not the same.

As can be seen, the fall-off of the density is 
in leading order proportional to $1/r^6$.
This was also obtained in \cite{visser-boul} but with 
a different factor, also in the correction terms. However, as can be noted in
(\ref{eq-8}), a singularity appears at the Schwarzschild radius. There the
vacuum fluctuations tend to infinity, rendering the semi-classical approach invalid. 
The qualitative conclusion drawn from this 
%%%new-start
is
%%%new-end
an
increase toward the central mass with some $1/r^n$
dependence. 
The origin of 
%%%new-start
%the
this coordinate 
%%%new-end
singularity at the Schwarzschild radius
is to neglect the back-reaction on the metric.

This observation leads to the proposal of the {\it pseudo-complex General
Relativity} (pc-GR) \cite{hess2009,schoenenbach2012}. In this modified theory
of General Relativity, an additional term on the right hand side of the Einstein
equations is required, which we associate with the energy-momentum tensor of the
vacuum fluctuations. For the energy-momentum tensor a model of an ideal
asymmetric fluid is applied. The advantage of this is that the back-reaction
of the vacuum fluctuations on the metric can be determined. However, due to its
phenomenological nature, an additional parameter appears. 
In its first version
\cite{schoenenbach2012} the density of the dark energy falls off as $1/r^5$ and
leads to $g_{00}=\left(1-\frac{2m}{r} + \frac{bm^3}{2r^3}\right)$.

Another interest is to investigate the vacuum fluctuations {\it within a star},
which requires the knowledge of the coupling of the dark energy to the mass 
distribution. In \cite{rodriquez2014a} a simple coupling model was applied in which the ratio of the dark energy density to the mass density is a constant.
In what follows we will 
%%%new-start
%deduce,
estimate, 
%%%new-end
using simple assumptions, 
the vacuum fluctuations as a function in the
radial distance, which will lead to a better proposal for the coupling, finally
applied in the next section.

For the ansatz of the mass distribution two models 
are considered: i) A constant mass
distribution within the star, which is the easiest to treat, 
and ii) a density which
simulates a realistic behavior. These distributions are implemented by hand and
not derived from the Tolman-Oppenheimer-Volkoff (TOV) equations, required by a 
consistent approach. We are, however, not interested in an exact description
but rather in obtaining an idea and motivation for a phenomenological ansatz of the
dark energy density as a function in $r$. This will then be applied in the next section.  

\subsection{A constant mass distribution}

The metric for a constant mass distribution $\varepsilon_m = \varepsilon_0$
is derived in \cite{adler}, i.e.,

\beqa
ds^2 & = & \left [ \frac{3}{2} \sqrt{1-\frac{R^2}{r_0^2}} - \frac{1}{2} \sqrt{1-\frac{r^2}{r_0^2}} \right ]^2 dt^2 - \frac{dr^2}{1-\frac{r^2}{r_0^2}} 
\nonumber \\
&&
- r^2\left ( d\vartheta^2 - \sin^2(\vartheta) d\varphi^2\right )
~~~,
\label{eq-9}
\eeqa
with $R$ as the radius of the star and $r_0^2=\frac{R^3}{2m}$. 

With (\ref{eq-6a}) and (\ref{eq-6b}), 
using the monopole approximation (\ref{eq-7}), 
we obtain for the expectation value of the 
$T_t^t$ and $T_r^r$ component 

\beqa
 \varepsilon_{\Lambda} & = & \frac{1}{4\pi r^2} \left \langle T_{~t}^{t~2D} \right \rangle 
~=~ - \frac{1}{216} \frac{\left ( \varepsilon_0 + 3p_r \right )^2}
{1 - \frac{8}{3} \varepsilon_0 r^2} 
\nonumber \\
  p_{r}^{\Lambda} & = & -\frac{1}{4\pi r^2} 
	\left \langle T_{~r}^{r~2D} \right \rangle 
	~=~  
	\frac{1}{72\pi r^2} \left ( 
  \varepsilon_0 + 3p_r^\prime r + 3p_r 
	\right.
	\nonumber \\
	&& \left.
	+ 2\pi 
	\frac{r^2 \left ( \varepsilon_0 - p_r \right )
	\left ( \varepsilon_0 + 3p_r \right ) }
  {1- \frac{8\pi}{3} \varepsilon_0 r^2}
  \right )
\label{eq-10}
\eeqa
The prime in $p_r^\prime$ of the matter pressure refers to a derivative in $r$.
We 
%%%new-start
%did not write down 
do not present
%%%new-end
the part for the matter density and the derivative
of 
%%%new-start
%the
its 
%%%new-end
pressure, because we are interested in the dark energy density part only.

Suppose that inside the star the fluid is 
%%%new-start
%isotropic. 
{\it isotropic}. 
%%%new-end
Then $p_r=p$ and using 
the TOV equations (see \cite{rodriquez2014a}) for the case of a constant mass density, we obtain \cite{adler}

\beqa
p_r =& \varepsilon_0 \frac{\sqrt{\frac{r_0^2 - r^2}{r_0^2- R^2}}- 1}
{3-\sqrt{\frac{r_0^2 - r^2}{r_0^2- R^2}}}
~~~.
\label{eq-11}
\eeqa

From now on, only $\varepsilon_\Lambda$ will be considered. 
Substituting (\ref{eq-11}) into
(\ref{eq-10}) leads for the dark energy density

\beqa
 \varepsilon_{\Lambda} & = & - \frac{\frac{1}{216} \varepsilon_0^2}{\left[ \frac{3}{2} \sqrt{1 - \frac{R^2}{r_0^2}} - \frac{1}{2} \sqrt{1 - \frac{r^2}{r_0^2}}\right]^2} 
~=~
- \frac{\frac{1}{576 \pi r_0^2} \varepsilon_0}
{\left[ \frac{3}{2} \sqrt{1 - \frac{R^2}{r_0^2}} - \frac{1}{2} \sqrt{1 - \frac{r^2}{r_0^2}}\right]^2}
~~~. 
\label{eq-12}
\eeqa
The factor in the numerator was rewritten as follows: With 
$m=\frac{4\pi}{3}\varepsilon_0 R^3$, resolving for $R^3$, 
we obtain with $r_0^2=\frac{R^2}{2m}$ 
(see the definition below (\ref{eq-9})) the value
$\varepsilon_0$ = $\frac{3}{8\pi r_0^2}$. In the factor
$\varepsilon_0^2$ in (\ref{eq-12}), this is 
substituted only for one $\varepsilon_0$.
%%%new-start
In this manner we show the explicit dependence of 
$\varepsilon_\Lambda$ to $\varepsilon_0$. 
%%%new-end

\begin{figure}[htp]
 \center
 \includegraphics[scale=1.0]{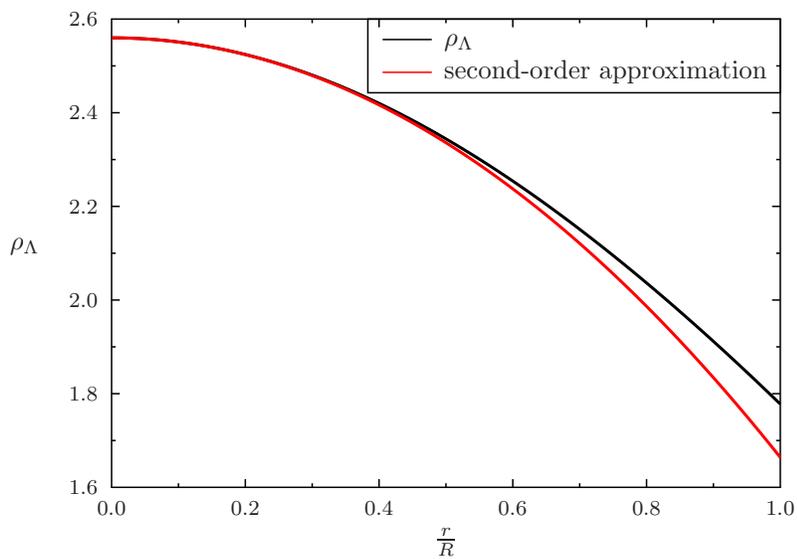}
 \caption{\label{fig-eps} 
The figure compares the radial dependance of the dark energy density  
for a star with constant matter density (see eq (\ref{eq-12})), 
radius $R$ = $\frac{32}{7}m \approx 4.6m$ and 
$\varepsilon_0 = 576 \pi r_0^2$ with  
its second-order approximation.
}
\end{figure}

The density, as given in (\ref{eq-12}) is expanded up to
leading order in $r^2$. In Fig. \ref{fig-eps} the 
quadratic approximation is compared to the exact relation
(\ref{eq-12}). Note that the major differences only occur near the surface. However, the approximation gets worse for lower
radii $R$.

In conclusion,
the dark energy density $\varepsilon_\Lambda$ 
is not only proportional to the constant mass density with a constant factor, but
there is an {\it additional $r$-dependence}, which can be approximated for $R$ not too small 
by a factor of the type  $\left(1-\beta r^2\right)$.
%is proportional to the constant mass density times a factor
%of the type $\left(1-\beta r^2\right)$, i.e., the relation is not a proportionality
%with a constant factor, but there is an {\it additional $r$-dependence}. 
The constant $\beta$ depends in (\ref{eq-12}) on the constant energy density and on the radius of the star. 
Assuming that the same relation of the coupling is valid 
%%%new-start
%also for a matter density which 
for a matter density which also 
%%%new-end
does depend on $r$, the coupling 
according to (\ref{eq-12}) should acquire an even more complicated $r$-dependence. However, in order to keep the matter simple, the
ansatz which depends on $r^2$, should suffice for a start. 

To resume, the considerations in this section suggests the following ansatz for the coupling of the dark energy to the
matter density:
 
\beqa
\varepsilon_\Lambda & = & \alpha \varepsilon_m 
\left[ 1 - \left( \frac{r}{r_0}\right)^2 \right]
~~~,
\label{eq-13}
\eeqa
where $\alpha$ is a factor as used in \cite{rodriquez2014a} and $r_0$ is an
additional parameter, {\it not to be confused} with the
parameter used in the model of a constant mass density.

\subsection{A non-constant mass distribution}

Let us assume that the mass density varies as

\beqa
\varepsilon_m & = & \varepsilon_0 \left[ 1-b_1 \left[\frac{r}{R}\right]^2\right]
~~~,
\label{eq-14}
\eeqa
which simulates realistic calculations \cite{rodriquez2014a}.
As a length element squared we use

\beqa
ds^2 & = & e^\nu dt^2 - e^\lambda dr^2 -r^2\left(d\vartheta^2
+ {\rm sin}^2\vartheta d\varphi^2\right)
~~~,
\label{eq-15}
\eeqa
where the metric for the interior of the star is 
proposed to behave as

\beqa
e^\nu & = & \left( 1 - \frac{r_s}{R}\right) 
e^{\frac{r_s}{2\left(R^3 - r_s R^2\right)\left(r^2 - r_0^2\right)}}
\nonumber \\
e^{-\lambda} & = & 1-\frac{8\pi}{3}\varrho_0 r^2\left( 1-b_1\frac{3r^2}{5R^2}\right)
~~~.
\label{eq-16}
\eeqa
In order that this connects smoothly to the Schwarzschild metric outside the star, 
the following condition has to be fulfilled:

\beqa
\frac{8\pi}{3} \varrho_0\left( 1- \frac{3}{5}b_1\right) & = & \frac{2m}{R^3}
\label{eq-16-a}
\eeqa

Following the same steps as described further above, we arrive for the
dark energy density at

\beqa
 \varepsilon_{\Lambda} = - \frac{m}{288\pi R^3} \frac{1 - \frac{3}{5} b_1}{\left ( 1 - \frac{2m}{R} \right )^2} 
\varepsilon_0 
\left [1 - \frac{2m r^2}{\left ( 1 -\frac{3}{5} b_1 \right )R^3}\left ( 1 - \frac{3}{5} b_1 \frac{r^2}{R^2} \right ) \right ]
~~~.
\label{eq-17}
\eeqa

The obtained $r$-dependence of the dark energy density on 
the matter density is now, as expected, more complicated.
Nevertheless, the main $(1-br^2)$-dependence is reproduced, 
with the correction factor proportional to $b_1\frac{r^2}{R^2}$.
Only near the surface ($r$ approximately $R$), this 
correction is large and may be attributed to
the ansatz of the matter density, i.e., it is expected to be
not very good there
%%%new-start
%.
%can be realized.
%%%new-end

Again, we arrive at 
the structure similar to the ansatz (\ref{eq-13}).
Thus, in the next section we use this ansatz to investigate if 
stars with larger masses than those obtained in \cite{rodriquez2014a}.

\section{Large mass stars}

The theory of neutron stars within the pseudo-complex General Relativity is published in 
\cite{rodriquez2014a}, assuming
a purely linear coupling between the dark energy density $\varepsilon_\Lambda$ and the corresponding quantity for the baryonic counterpart $\varepsilon_m$:    

\begin{equation}
\label{eq:PurLinCoupling}
\varepsilon_\Lambda=\alpha \varepsilon_m
\end{equation}
The coupling parameter was chosen to lie in the range $(-1,0)$ and under such conditions, stable solutions were obtained having masses for the baryonic component 
of up to $[6-7]M_\odot$. In the present work, we repeat such calculations using instead of (\ref{eq:PurLinCoupling}) 
the new coupling relation 
(\ref{eq-13}).

\begin{figure}[htp]
 \center
 \includegraphics[scale=0.6]{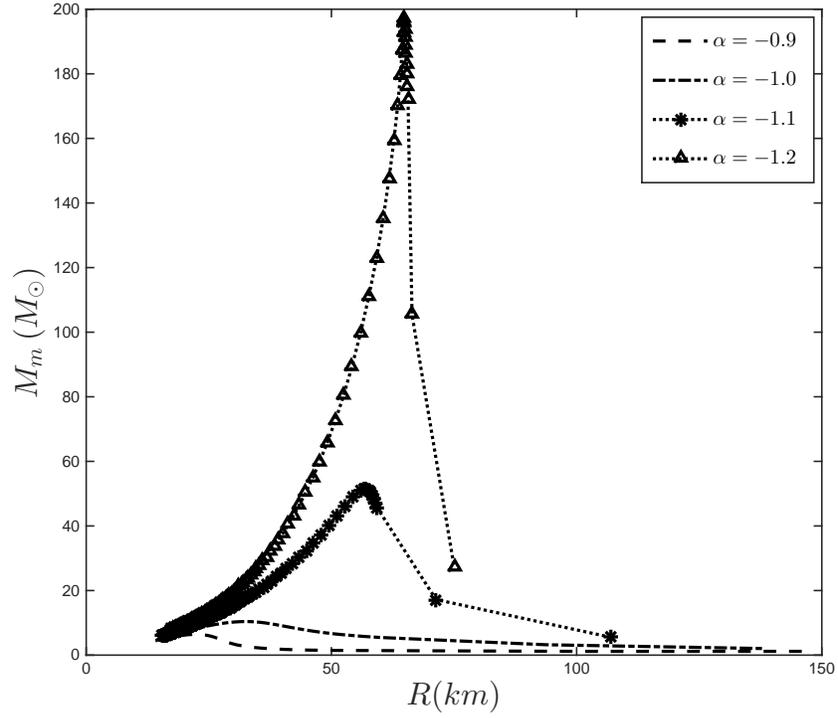}
 \caption{\label{Figure1} Mass of the baryonic component vs. radius for different values of the coefficient $\alpha$. The central $\Lambda$-term pressure $p_{\Lambda c}$ has been fixed to $P_{\varepsilon_{mc}=10^{-2}\varepsilon_0}$, the $r_0$ parameter to $100\;km$ and the central baryonic energy density $\varepsilon_{mc}$ between $[4\cdot10^{-2}-10]\varepsilon_0$.}
\end{figure} 

As can be appreciated from Fig. \ref{Figure1}, the new coupling relation allows the study of solutions with values 
of $\alpha < -1$ whose masses can acquire considerable high values of up to 200 solar masses, in comparison with the previous work. This can be understood as a consequence of the stronger fall-off of the new relation toward the radius of the star that causes the dark energy repulsion near the surface to be weaker than the pure linear model (\ref{eq:PurLinCoupling}), while it remains strong near the center. Thus, larger stellar masses can be maintained without collapsing and 
the repulsion is not strong enough to evaporate the surface.

\begin{figure}[htp]
 \center
 \includegraphics[scale=0.6]{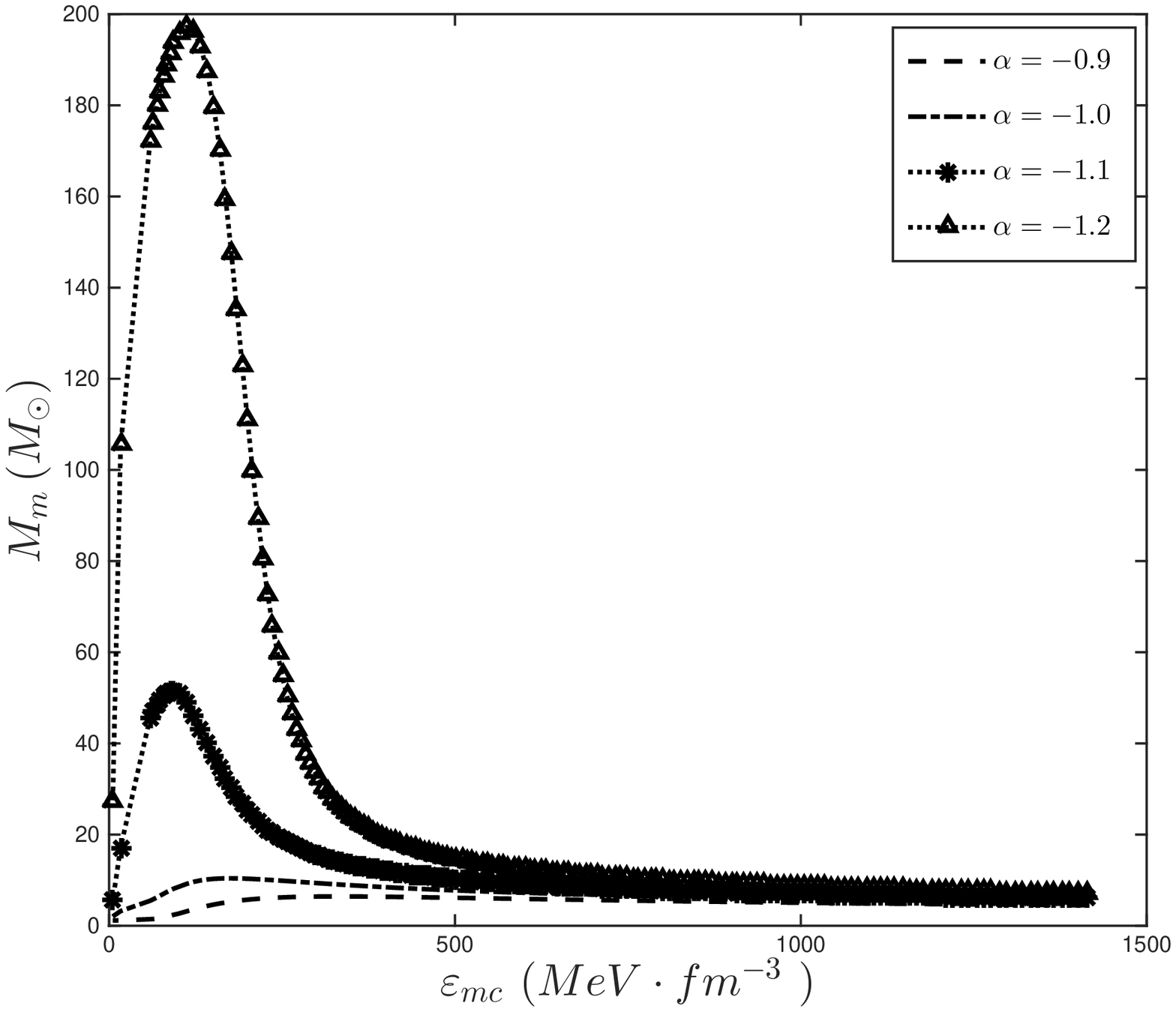}
 \caption{\label{Figure2} Mass of the baryonic component vs. its central energy density for different values of the coefficient $\alpha$. The central $\Lambda$-term pressure $p_{\Lambda c}$ has been fixed to $P_{\varepsilon_{mc}=10^{-2}\varepsilon_0}$, the $r_0$ parameter to $100\;km$ and the central baryonic energy density $\varepsilon_{mc}$ between $[4\cdot10^{-2}-10]\varepsilon_0$.
%%%new-start
Stability can be proven, as explained in 
\cite{rodriquez2014a}, investigating how the curve bends for large central density. However, we could not proceed because
data on the baryonic properties were not available for that region.
%%%new-end
}
\end{figure} 
 
The presence of these more massive solutions can be 
appreciated in Fig. \ref{Figure2} where already one can identify those which have $\frac{dM}{d\varepsilon_c}<0$ and therefore are unstable against small perturbations. Unfortunately, the stability of the solutions with $\frac{dM}{d\varepsilon_c}>0$ cannot be completely assured by simple criteria as those employed in \cite{rodriquez2014a} and a full perturbative method needs to be carried out. 
%%%new-start
%However, numerical 
Numerical 
%%%new-end
instabilities inhibited a more 
profound study.
Checking the previous, necessary condition however tells 
us that at least such massive objects could 
%%%new-start
%in principle exist.
{\it in principle exist}.
%%%new-end

\begin{figure}[htp]
 \center
 \includegraphics[scale=0.6]{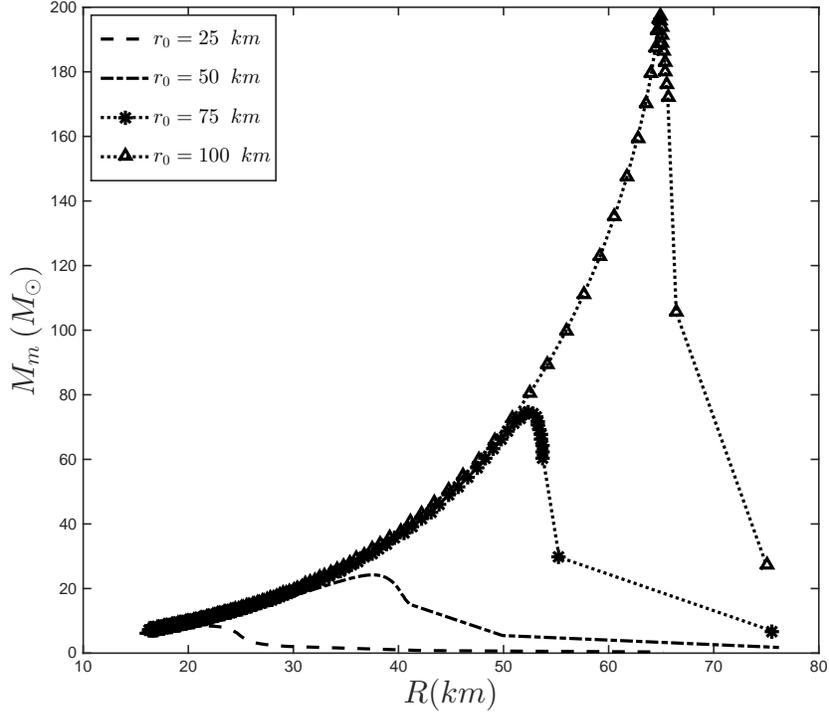}
 \caption{\label{Figure3} Mass of the baryonic component vs. radius for different values of the parameter $r_0$. The central $\Lambda$-term pressure $p_{\Lambda c}$ has been fixed to $P_{\varepsilon_{mc}=10^{-2}\varepsilon_0}$, the coupling parameter $\alpha$ to $-1.2$ and the central baryonic energy density $\varepsilon_{mc}$ between $[4\cdot10^{-2}-10]\varepsilon_0$.}
\end{figure}

\begin{figure}[htp]
 \center
 \includegraphics[scale=0.6]{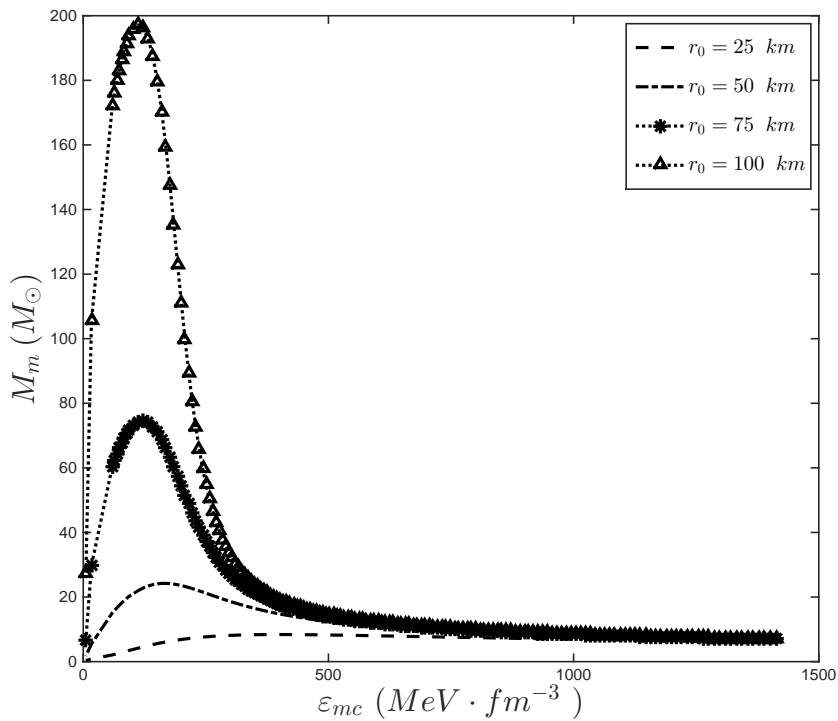}
 \caption{\label{Figure4} Mass of the baryonic component vs. its central energy density for different values of the parameter $r_0$. The central $\Lambda$-term pressure $p_{\Lambda c}$ has been fixed to $P_{\varepsilon_{mc}=10^{-2}\varepsilon_0}$, the coupling parameter $\alpha$ to $-1.2$ and the central baryonic energy density $\varepsilon_{mc}$ between $[4\cdot10^{-2}-10]\varepsilon_0$.}
\end{figure}

Within the current study, an additional parameter $r_0$ is introduced. As can be seen in 
Figs. \ref{Figure3} and \ref{Figure4}, larger values of $r_0$ translate into larger values of the mass corresponding to the baryonic component. The value of $r_0$ cannot be {\it a priori} determined and therefore four values covering the whole range in which the radii of the solutions lie\footnote{Please note that the value chosen for $r_0$ will modify the stellar radius.} have been chosen.

\section{Conclusions}

Using semi-classical Quantum Mechanics,
the mass-induced vacuum fluctuations outside and inside a mass distribution were
determined within a monopole approximation. Outside a mass distribution the
dominant part of the dark energy density falls off as $\frac{1}{r^6}$, which is
in qualitative  
agreement to calculations in four dimensions 
\cite{visser-boul}. This
finding justifies the ansatz for the dark energy density, as used in the pc-GR  \cite{hess2009,schoenenbach2012}.

Inside the mass distribution similar calculations lead to a deviation of
the linear coupling, as assumed in \cite{rodriquez2014a}. The encountered dependence
of the coupling between the dark energy density and the mass density includes
an extra factor, depending on the radial distance $r$. This finding permits a 
new,
%%%new-start
phenomenological
%%%new-end
ansatz for the coupling, which allows a large dark energy density in the center
of a star, therefore sustaining a large mass, and a low density near the surface, thus not evaporating the outer layers of the star, as found in \cite{rodriquez2014a}.

The discussion presented is a {\it prove of principle} that arbitrary large masses 
of stars are possible, 
which indicates that the so called black holes may be
in fact large 
%%%new-start
%neutron 
%%%new-end
stars, though due to the strong gravitational field the physics
at the surface and in its interior 
is  expected to change significantly,
%%%new-start
compared to known neutron stars.
%%%new-end

\section{Acknowledgement}
{\textsc{Peter O. Hess}} acknowledges the financial support from DGAPA-PAPIIT (IN100315).
{\textsc{Isaac Rodr\'{i}guez}} acknowledges the financial support from the {\it Alexander von Humboldt Stiftung} and FIAS. 
Gunther Caspar acknowledges financial support from FIAS.
\label{sec:Ack}

\end{document}